# Phonons reflect dynamic spin-state order in $LaCoO_3$

Alsu Ivashko,[1] Taishun Manjo,[2,3] Maximilian Kauth,[1] Yuliia Tymoshenko,[1] Adrian M. Merritt,[4] Klaus-Peter Bohnen,[1] Rolf Heid,[1] Michael Merz,[1,5] Andreas Eich,[1] John-Paul Castellan,[1,6] Alexandre Ivanov,[7] Nathaniel Schreiber,[8] Hong Zheng,[8] J. F. Mitchell,[8] Martin Meven,[9,10] Jitae T. Park,[11] Daisuke Ishikawa,[2,3] Yuiga Nakamura,[12] Alfred Q. Baron,[2,3] and Frank Weber[1,†]

[1]Institute for Quantum Materials and Technologies, Karlsruhe Institute of Technology, 76131 Karlsruhe, Germany
[2]Materials Dynamics Laboratory, RIKEN SPring-8 Center, RIKEN, 1-1-1 Kouto, Sayo, Hyogo 679-5148 Japan
[3]Precision Spectroscopy Division SPring-8/JASRI, 1-1-1 Kouto, Sayo, Hyogo 679-5198 Japan
[4]Physikalisches Institut, Karlsruhe Institute of Technology, 76021 Karlsruhe, Germany
[5]Karlsruhe Nano Micro Facility (KNMFi), Karlsruhe Institute of Technology, 76131 Karlsruhe, Germany
[6]Laboratoire Léon Brillouin (CEA-CNRS), CEA Saclay, F-91911 Gif-sur-Yvette, France.
[7]Institut Laue-Langevin, 71 avenue des Martyrs CS 20156, 38042 Grenoble Cedex 9, France.
[8]Materials Science Division, Argonne National Laboratory, Lemont, IL 60439, USA.
[9]Institute of Crystallography, RWTH Aachen, University, 52056 Aachen 52056, Germany
[10]Jülich Centre for Neutron Science (JCNS), Forschungszentrum Jülich GmbH at Heinz Maier-Leibnitz Zentrum (MLZ), 85748 Garching 85748, Germany
[11]Heinz Maier-Leibnitz Zentrum (MLZ), Technische Universität München, 85748 Garching, Germany
[12]Diffraction and Scattering Division, SPring-8/JASRI, 1-1-1 Kouto, Sayo, Hyogo 679-5198 Japan

We investigate lattice dynamics in $LaCoO_3$ using inelastic neutron and x-ray scattering over T = 2–650 K, spanning the spin-state crossover at $T_1 \approx 100$ K and the insulator–metal transition at $T_2 \approx 550$ K. Comparison with quasi-harmonic ab-initio lattice-dynamical calculations helps reveal anomalous softening of a ≈10-meV oxygen phonon, confined to the temperature interval $T_1 \leq T \leq T_2$ and localized in momentum space at $\mathbf{q}_{SSO} = (½,½,½)_c$. This wave vector corresponds to the spin-state ordering originally proposed by Goodenough [*J. Phys. Chem. Solids* **6**, 287-297 (1958)]. Our results therefore provide momentum-resolved evidence for dynamic correlations of high-spin and low-spin $Co^{3+}$ states in $LaCoO_3$, linking spin-state fluctuations to anomalous phonon renormalization.

Introduction

Spin-crossover (SCO) compounds—ranging from molecular Fe(II) complexes to solids such as cobaltates and ferropericlase—exhibit reversible changes in electronic spin state that can be driven by temperature, pressure, or light. These materials are central to fundamental research on coupled spin–lattice degrees of freedom[1-3], properties of the Earth's lower mantle[4-6] and emerging applications[7,8] exploiting bistability, large entropy changes, and tunable mechanical and optical properties. While the structural change induced by the spin transition can be subtle, the crossover can create strong changes in local bonding leading to changes in the atomic motions. Thus measurements of phonon dispersion can be a sensitive probe of SCO materials.

Among oxide materials, $LaCoO_3$ is a canonical example of a thermally driven SCO system: with increasing temperature, $Co^{3+}$ ions evolve from a low-spin to higher-spin configuration, producing marked anomalies in thermal expansion[9,10], elastic moduli[11], and heat capacity[12]. Despite extensive study[13-20], the magnetic properties of $LaCoO_3$ and the temperature evolution of the $Co^{3+}$ spin state remain debated.

At low temperature $LaCoO_3$ adopts a slight rhombohedral distortion from the cubic symmetry of a simple perovskite[a] [Fig. 1(a)] and magnetic susceptibility measurements [Fig. 1(b)][21] reveal two crossover regions: a sharp one near $T_1 \approx 100$ K and a broader one near $T_2 \approx 550$ K. These have been linked to Co-localized spin-state transitions of $Co^{3+}$ ($3d^6$) from the low-spin (LS, $t_2g^6$, S = 0) ground state to excited intermediate-spin (IS, $t_2g^5e\,g^1$, S = 1) and high-spin (HS, $t_2g^4e\,g^2$, S = 2) states. Goodenough initially proposed[14] that, at intermediate temperatures ($T_1 \leq T \leq T_2$), where LS and HS populations are comparable, the different ionic radii of LS and HS states of $Co^{3+}$ ($r_{LS}/r_{HS} \approx 0.9$) yield a spin-state order (SSO) consisting of sets of alternating (111)-planes, in one of which the Co ions are predominantly in a HS state and in the other in a LS state. The corresponding ordering wave vector is $\mathbf{q}_{SSO} = (½,½,½)_c$. Although diffraction experiments found no evidence of such order[22,23], indirect probes suggest LS–HS coexistence in this regime[15,17,24]. In 1995, LDA+U calculations by Korotin et al.[25] suggested instead a substantial IS population, implying orbital order of Co $e_g$ orbitals with an ordering vector $\mathbf{q}_{OO}$

[a] Throughout the manuscript, we label wave vectors with indices "c" and "rh" if they are given in the pseudo-cubic or rhombohedral unit cell, respectively. For more details please see SI.

= (½,½,0)$_c$. This IS-state scenario has since been invoked regularly to interpret experimental data[18,26,27].

Here, we present an investigation of lattice dynamics in $LaCoO_3$ across the temperature-driven SCO employing inelastic x-ray (IXS) and neutron scattering (INS) and extending beyond previous studies[28,29]. We use the measured phonon dispersion near the wave vectors $\mathbf{q}_{SSO}$ and $\mathbf{q}_{OO}$ as a dynamic probe of the SCO transition. We also experimentally verify the rhombohedral disortion by X-ray (XRD) and neutron diffraction (ND), using *ab-initio* lattice dynamics based on density functional perturbation theory (DFPT). Quasi-harmonic (QH) calculations, incorporating the reported temperature dependence of lattice constants[23], predict conventional phonon softening due to thermal expansion and, thus, help us to identify anomalous softening because of the SCO. Also, two-phonon calculations[30,31] help explain the loss of spectral contrast at elevated temperatures. Our main finding is softening of an oxygen-dominated low-energy phonon confined to $\mathbf{q}_{SSO}$ = (½,½,½)$_c$ present only for temperatures between the spin-state and insulator-metal crossover, $T_1 \leq T \leq T_2$. This provides momentum-resolved evidence for dynamic $Co^{3+}$ SSO in $LaCoO_3$ of the kind proposed by Goodenough.

## Methods

Electronic and lattice-dynamical properties were calculated using density-functional theory (DFT) and DFPT within the mixed-basis pseudopotential (MBPP) framework[32,33]. QH calculations, based on experimentally determined lattice constants[23], were used to capture phonon frequency shifts upon heating (see SI). A single crystal was grown by the floating-zone method, yielding an ≈8 cm rod. One half was used for INS, and parts from the other for magnetization and IXS.

XRD was carried out at BL02B1[34] (SPring-8, Japan), and complementary neutron diffraction at the HEiDi diffractometer[35] (MLZ, Germany). Results of the crystallographic refinements and some of the data used to create Figures 1-5 can be accessed online[36]. Lattice dynamics were studied with meV-resolution IXS at BL43LXU[37,38] (SPring-8) using 21.747 keV photons (Si(11 11 11) backscattering, ~1.4 meV resolution) and a two-dimensional analyzer array enabling simultaneous access to multiple momentum transfers (see [38] for more details of the current setup). Neutron triple-axis spectrometry (TAS) was performed on 1T (LLB), IN8[39,40] (ILL) and PUMA[41] (MLZ) with fixed final energy of 14.7 meV and PG filtering. The crystal was aligned in the [110]$_c$–[001]$_c$ plane and mounted in closed-cycle or liquid-helium cryofurnaces, enabling measurements up to 650 K. Energy scans were performed at constant **Q** = **τ** + **q**, with **τ** a reciprocal lattice vector and **q** the phonon wave vector in the first Brillouin zone. Wave vectors are given in pseudo-cubic ("c") or rhombohedral ("rh") notation as indicated.

Results

$LaCoO_3$ features a rhombohedrally distorted perovskite structure ( $R\bar{3}c$ , #167) for temperatures below 1600 K.[42] Several neutron and x-ray diffraction studies reported no further structural changes on cooling[18,22,23]. One report however suggested a monoclinic distortion[27] ($I2/a$, #15) claiming that previous powder diffraction experiments could have easily missed it. Subsequent analysis of neutron powder diffraction[43] did not confirm a monoclinic distortion and set a small upper boundary. To clarify this question, we performed synchrotron-based state-of-the-art single crystal XRD (T = 150K – 400K), and complementary single-crystal neutron diffraction (T = 2K – 600K). Here, we report that all of our data can be well described in the rhombohedral structure ($R\bar{3}c$, #167) [see Supplemental Information (SI) for more details]. The analysis of XRD data excludes a monoclinic distortion in our sample as is corroborated by neutron diffraction revealing a normal behavior of the oxygen thermal atomic displacement parameters with temperature (see SI).

Lattice-dynamical calculations are essential for $LaCoO_3$, where the rhombohedral lattice distortion generates a highly complex phonon spectrum with up to 60 branches, *e.g.*, along $[111]_c$ which corresponds simultaneously to $[111]_{rh}$ and $[100]_{rh}$ in a multi-domain sample [Fig. 1(c)]. Our ab initio approach reproduces not only dispersions but also momentum- and energy-resolved phonon intensities, enabling direct comparison with IXS spectra. At the R point, **Q** = $(2.5,2.5,2.5)_c$, the measured spectrum agrees with the calculated one-phonon contribution, provided a two-phonon background[31] is included [Fig. 2(a)]. We found that the small x-ray beam (~50 μm) probes a single domain, reducing the problem to one rhombohedral vector, **Q** = $(5,5,5)_{rh}$, where only four phonons carry finite intensity—greatly simplifying the analysis. Measurements centered at **Q** = $(2.5,2.5,2.5)_c$ and **Q** = $(0.5,3.5,3.5)_c$ [Figs. 2(b,c)] confirm this picture. For instance, the high-energy mode near 70 meV is absent at the former and present at the latter wave vector, in agreement with single domain calculations. These high-energy Co-O breathing-type modes, extensively studied in cuprates[44-47] and manganites[48,49], show no anomalous behavior at finite momentum transfers except for a general suppression of spectral weight on heating (see SI). Phonons at intermediate energies of 40-45 meV, however, display an intriguing opposite trend in their temperature dependent energies (see SI) which we will further investigate and report on at a later time. Here, we focus on an anomalous softening of a ≈10-meV phonon observed at the cubic R point. Its displacement pattern is dominated by oxygen motions (see below) and, therefore, we investigated this phonon by INS because of the mass dependence of x-ray scattering intensities. At **Q** = $(1.5,1.5,1.5)_c$ and $(1.5,1.5,2.5)_c$ [Figs. 3(a,b)], the 10-meV mode[b] exhibits strong softening with increasing temperature and linewidth broadening near $T_1$, while no anomaly is found for the low-energy phonon branch dispersing at **Q** = $(1.6,1.6,1.6)_c$ [Fig. 3(c)] or in higher-energy modes at the R point [Fig. 3(b)]. Similarly, the lowest-energy modes at the M point, **Q** = $(1.5,2.5,0)_c$, shows no particularly strong softening nor broadening [Fig. 3(d)]. QH calculations, which reproduce most phonon energies, fail to capture this effect [Figs. 4(a,b)]: the 10-meV mode at the R point softens well below the QH prediction between $T_1 \approx 100$ K and $T_2 \approx 550$ K, before converging with theory at higher temperatures (red symbols). This anomalous renormalization is absent at **Q** = $(1.6,1.6,1.6)_c$ [solid black symbols in Fig. 4(a)], in the higher-energy mode at the R point [open black symbols in Fig. 4(b)] as well as at the M point [solid blue symbols in Fig. 4(b)], confirming its restriction to the oxygen phonon at the R point. We further observe that the experimentally observed linewidths of the 10-meV modes at the R [red symbols in Fig. 4(c)] and M points [blue symbols in Fig. 4(c)] are very similar at low temperatures but the R point mode acquires an additional broadening just below the crossover temperature $T_1$ not observed for the M point mode.

The temperature dependence of the anomaly closely parallels that of (para-)magnetic scattering from polarized neutron scattering[50] (Fig. 5), indicating a common origin, i.e., the spin-state transitions of $Co^{3+}$.

<u>Discussion</u>

Our results agree with previous phonon investigations as far as similar regions of momentum and energy space as well as temperature were probed. However, most Raman studies did not go beyond room temperature[51,52], similar to a neutron scattering study which was limited to 200 K.[28] IXS studies[29,53] exclude the (rhombohedral) zone center and, thereby, the pseudo-cubic R point. In fact, the 10-meV mode at the M point was investigated by IXS[53] and was found to feature no anomalous behavior, in agreement with our findings [Figs. 3(d), 4(b,c)]. Apart from these limitations, all the previous studies lacked detailed lattice dynamical calculations. In particular, QH modelling is an indispensable tool[54].

---

[b] The scattering intensities near 10 meV measured at ***Q*** = $(1.5,1.5,1.5)_c$ and ***Q*** = $(1.5,1.5,2.5)_c$ (both corresponding to rhombohedral zone-center wave vectors) represent the same phonon mode.

Because the SCO alters Co–O bonding and octahedral stiffness, comparison of our results with spin-state models in $LaCoO_3$ is revealing. The LS–HS scenario, originally proposed by Goodenough[14,22,55], predicts a SSO consisting of sets of alternating (111)-planes, in one of which the Co ions are predominantly in a HS state and in the other in a LS state. The different ionic radii of $Co^{3+}$ in the HS and LS states increase and decrease, respectively, the size of the $CoO_6$ octahedron yielding structural correlations with a periodicity of $\mathbf{q}_{SSO} = (\frac{1}{2},\frac{1}{2},\frac{1}{2})_c$, consistent with our observation of phonon renormalization at the R point. Such correlated lattice fluctuations are expected to be strongest between $T_1$ and $T_2$, in agreement with the observed anomaly. Moreover, calculations identify the 10-meV mode as a pure oxygen vibration involving tilted rotations of the $CoO_6$ octahedra (inset, Fig. 5) rationalizing its high sensitivity to dynamic correlations of oxygen displacements. In contrast, the IS model invokes orbital order doubling the pseudo-cubic unit cell only along two principal directions (see Fig. 6 in [25]). The corresponding lattice distortions are expected to produce correlations at $\mathbf{q}_{OO} = (\frac{1}{2},\frac{1}{2},0)_c$, which could be accommodated in a monoclinic structure[27]. However, XRD shows that $LaCoO_3$ is rhombohedral and, correspondingly, we find no phonon anomalies of the 10-meV mode at $(\frac{1}{2},\frac{1}{2},0)_c$ [see Figs. 3,4, SI and [53]]. Our results therefore favor LS–HS-type fluctuations over an IS-driven mechanism, in line with several recent reports[17,19,20]. Notably, soft x-ray scattering investigations have also excluded the presence of $Co^{3+}$ IS sites in thin films of $LaCoO_3$ (Ref. [56]). The insights presented here into the dynamics of $LaCoO_3$ demonstrate that phonons are sensitive to the periodicity of lattice correlations when the latter are dynamic and below the detection limit of diffraction[57,58]. Individual phonon oscillation patterns can provide enhanced sensitivity and promote element-selective renormalization effects.

Summary

In summary, we establish an anomaly in an oxygen-dominated low-energy phonon in $LaCoO_3$ confined to the R point and present only between the spin-state and insulator-metal crossovers. This provides momentum-resolved evidence for dynamic $Co^{3+}$ LS–HS correlations, supporting the scenario of a dynamic LS-HS spin-state order at intermediate temperatures in $LaCoO_3$ of a kind with that proposed by Goodenough[14,22,55].

**Acknowledgements**

We wish to thank D. Phelan, S. Rosenkranz and R. Osborn for helpful discussions. The work of Y. Tymoshenko was supported by the consortium DAPHNE4NFDI in the context of the work of the NFDI e.V.. The consortium is funded by the Deutsche Forschungsgemeinschaft (DFG, German Research Foundation) - project number 460248799. The work of A. Merritt was supported by BMBF, Germany, in the framework of ErUM-Pro (Project No. 05K22VK1, NMO4PUMA). Work in the Materials Science Division of Argonne National Laboratory (crystal growth, magnetic characterization) was supported by the U.S. Department of Energy, Office of Science, Basic Energy Sciences, Materials Science and Engineering Division. Parts of the presented data were measured on the single crystal diffractometer HEiDi jointly operated by RWTH Aachen University and JCNS/Forschungszentrum Jülich at the MLZ within the JARA initiative. The IXS experiments were performed at the RIKEN Quantum NanoDynamics Beamline, BL43LXU, with approval of RIKEN SPring-8 Center (Proposal No. 20220075). Neutron scattering beam time at the ILL was provided under the proposal number 7-01-586 [40].

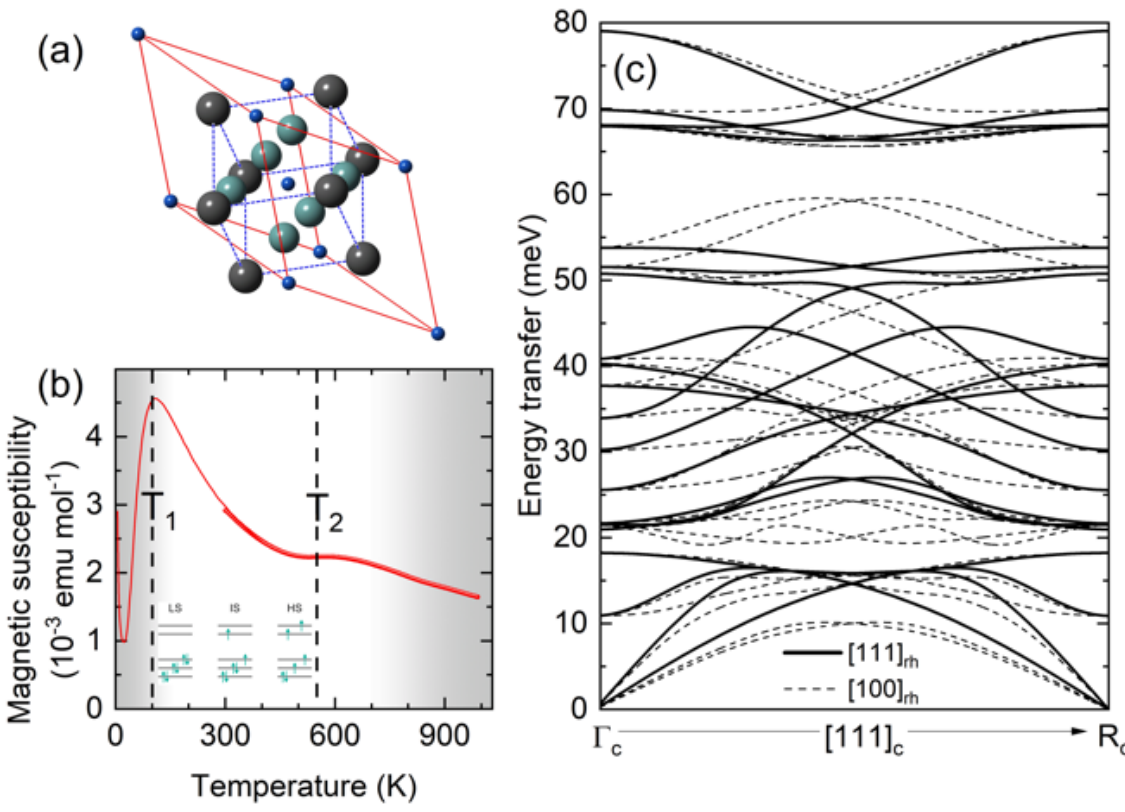

**FIG. 1**. (a) Pseudo-cubic perovskite structure of $LaCoO_3$. The red solid lines indicate the orientation of the low-temperature rhombohedral unit cell used for the phonon calculations. (b) Magnetic susceptibility of $LaCoO_3$. Data at $T \geq 300$ K were measured on our sample. Data below room temperature are taken from Ref. [21]. Dashed vertical lines denote the crossover temperatures $T_1$ and $T_2$ (see text). The inset at the bottom depicts the configuration of the 3*d* electrons of the $Co^{3+}$ ion in the low- (LS), intermediate- (IS), and high-spin (HS) state. (c) Calculated phonon dispersion shown for the pseudo-cubic $[111]_c$ direction, which corresponds to two rhombohedral directions, i.e., $[111]_{rh}$ (solid lines) and $[100]_{rh}$ in multi-domain samples (dashed lines).

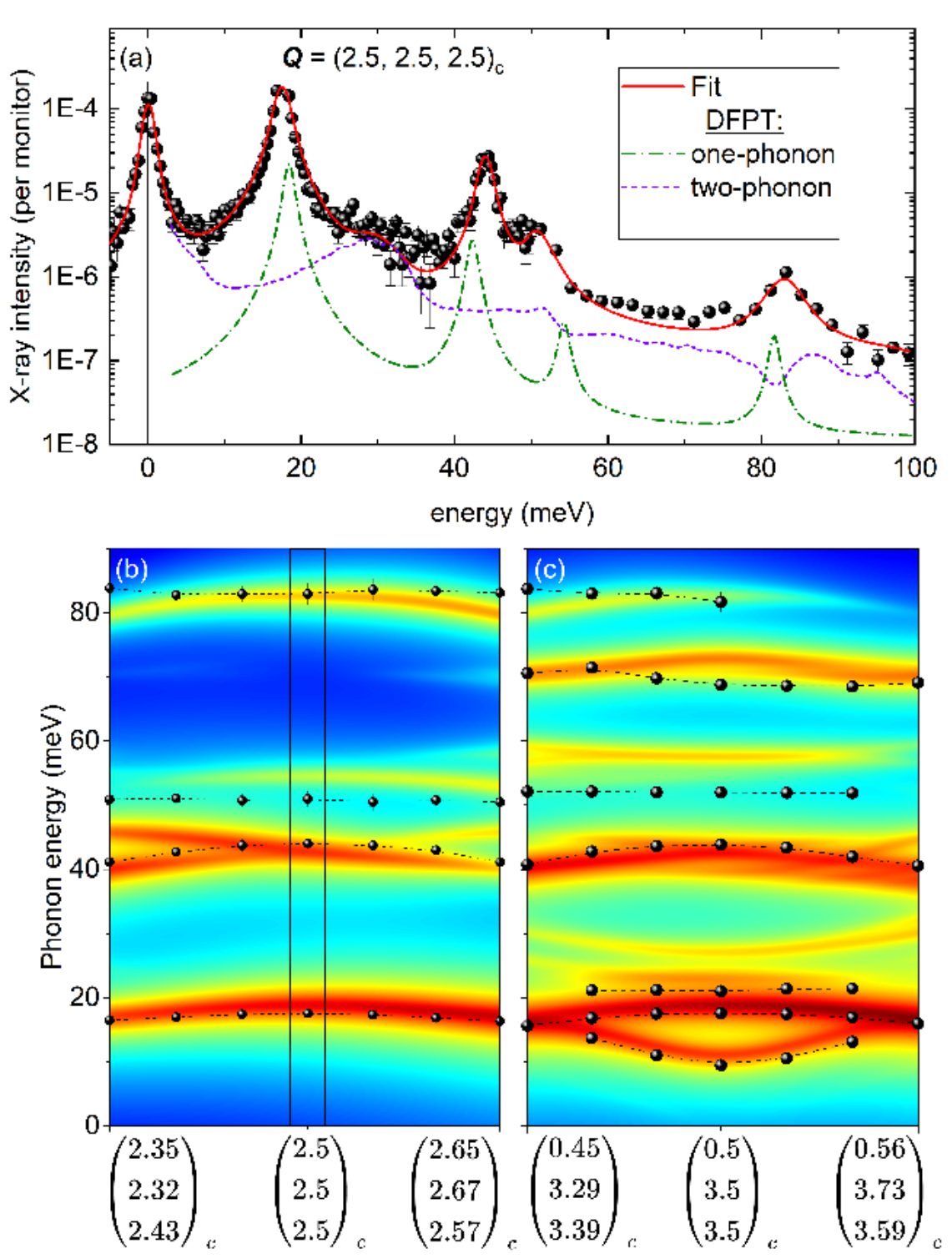

**FIG. 2**. (a) Inelastic x-ray scattering data taken at T = 100 K and ***Q*** = $(2.5,2.5,2.5)_c$ [≡ $(5,5,5)_{rh}$]. The fit (red, solid line) consists of DHO peaks for one-phonon excitations (convoluted with the experimental resolution) on top of a two-phonon background (purple, short-dashed) estimated from our DFPT calculations and the code presented in [31]. Calculated one-phonon intensities are shown as green dash-dotted line (includes a constant background of $10^{-8}$ for visibility) and represent single-domain spectra for the above given rhombohedral wave vector ***Q*** = $(5,5,5)_{rh}$ (see text for more details). We note the relative intensity of the 1- and 2- phonon calculations is *not* free, so the agreement is excellent. (b)(c) Color-coded calculated intensities (log. Scale, red: high, blue: low) for wave vectors probed by the horizontal row of seven IXS analyzers/detectors when the main analyzer/detector is set to (b) ***Q*** = $(2.5,2.5,2.5)_c$ and (c) $(0.5,3.5,3.5)_c$ (see [38] for more details of the setup). Symbols denote observed phonon energies. The box in (b) marks the position for which data are shown in panel (a). Calculations presented in this figure used energies scaled by a factor of 1.03 providing the best average agreement with the experimental results (see text).

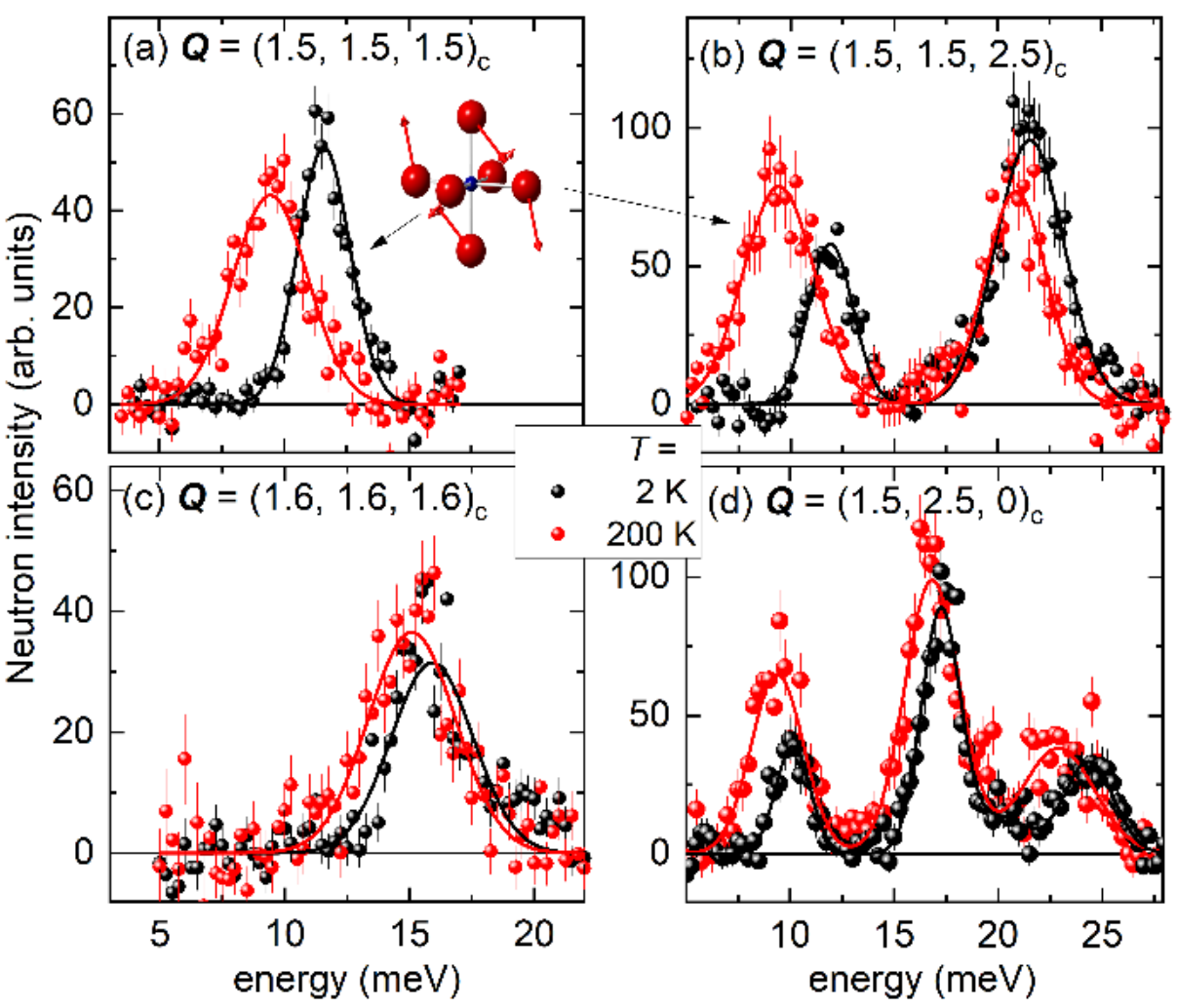

**FIG. 3**. Inelastic neutron scattering data at two R points, (a) **Q** = $(1.5,1.5,1.5)_c$, and (b) **Q** = $(1.5,1.5,2.5)_c$, (c) slightly off the R point at **Q** = $(1.6,1.6,1.6)_c$ and (d) the M point at **Q** = $(1.5,2.5,0)_c$ for temperatures T = 2 K (black symbols) and 200 K (red symbols). A linear background has been subtracted. Lines denote Gaussian fits. The inset in panel (a) shows the calculated displacement pattern of the 10-meV mode at the R point. Note that the scattering intensities near 10 meV measured at **Q** = $(1.5,1.5,1.5)_c$ and **Q** = $(1.5,1.5,2.5)_c$ represent the same phonon mode.

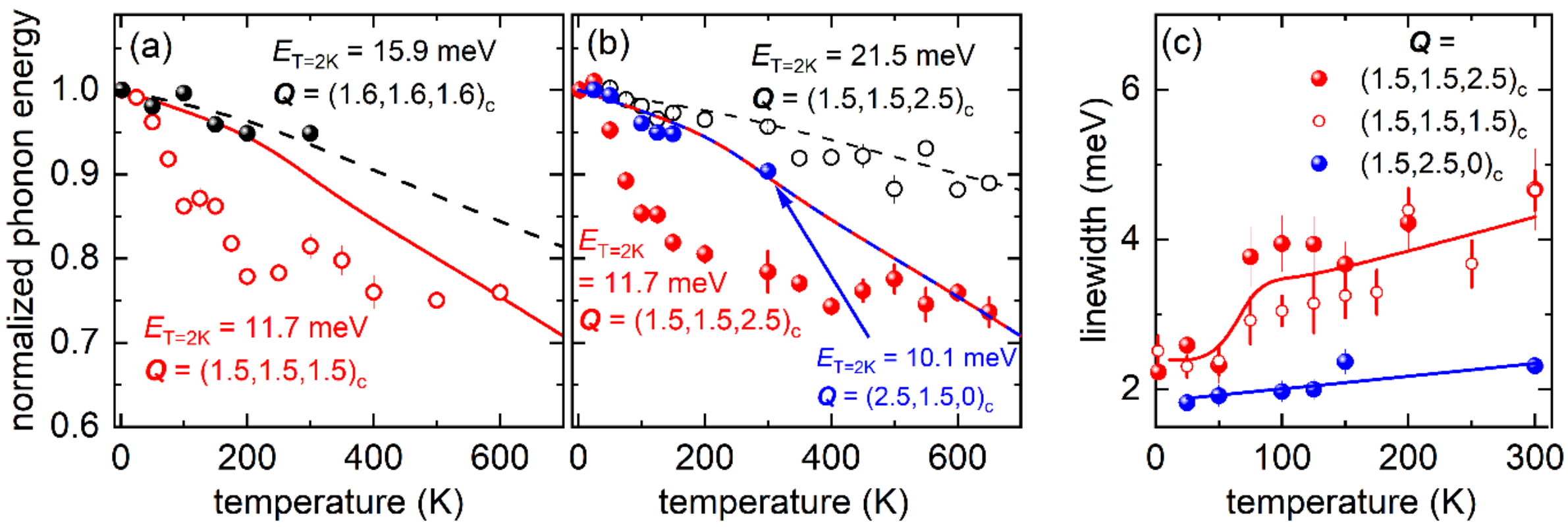

**FIG. 4**. (a,b) Temperature dependent energies, normalized at T = 2K, of phonon peaks (symbols) observed at (a) **Q** = $(1.5,1.5,2.5)_c$ (red symbols), **Q** = $(1.6,1.6,1.6)_c$ (black symbols) and (b) **Q** = $(1.5,1.5,2.5)_c$ (red & black symbols), **Q** = $(1.5,2.5,0)_c$ (blue symbols). Color-coded lines in (a) and (b) are predictions for the expected phonon softening due to thermal expansion from quasi-harmonic calculations (see text). (c) Experimentally observed linewidths of the lowest-energy phonon modes at **Q** = $(1.5,1.5,2.5)_c$ (red dots), **Q** = $(1.5,1.5,2.5)_c$ (red circles) and **Q** = $(1.5,2.5,0)_c$ (blue dots). Color-coded lines in (c) are guides to the eye.

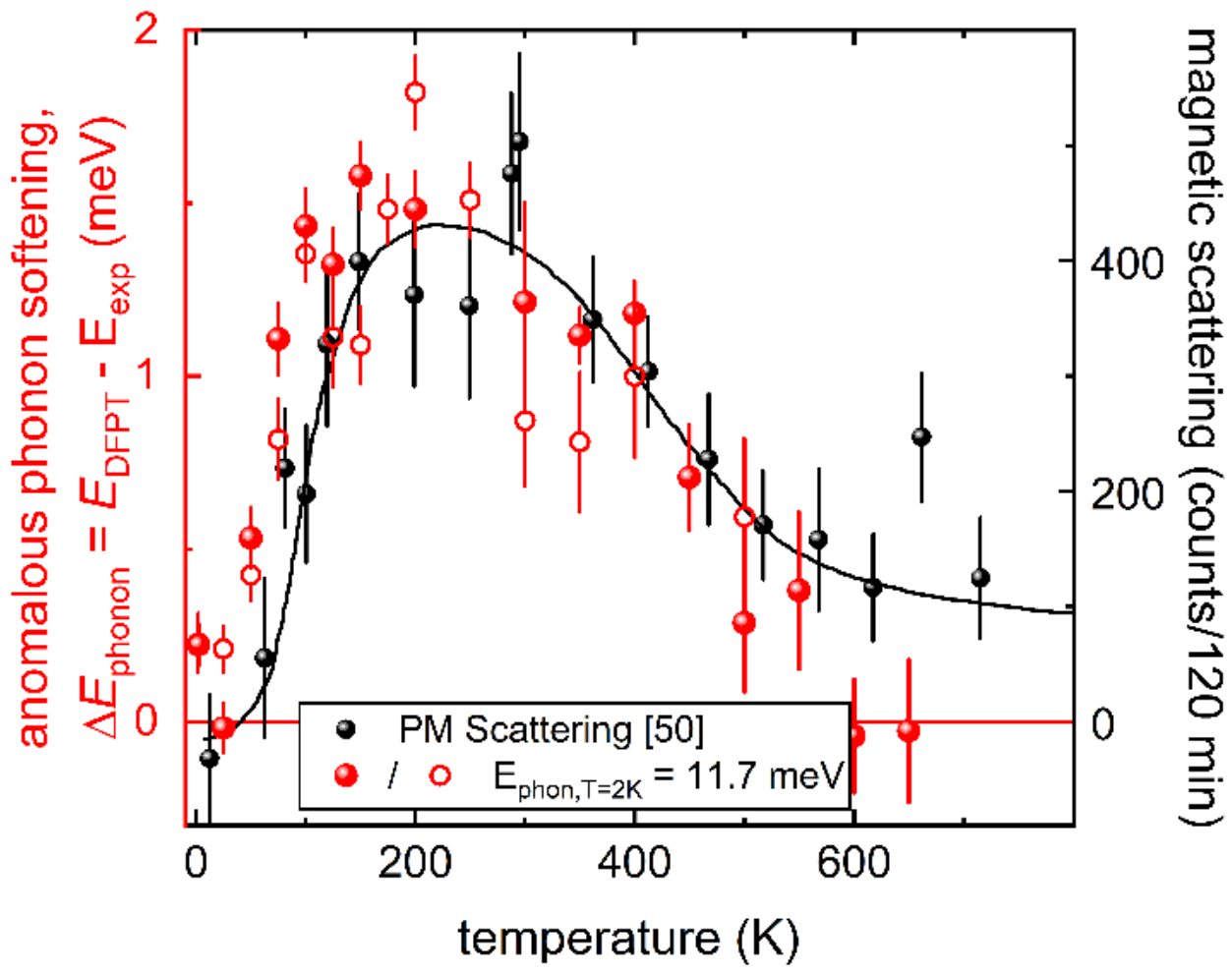

**FIG. 5**. Temperature dependence of the anomalous softening of the 10 meV phonon mode observed at the R point, i.e., **Q** = $(1.5,1.5,1.5)_c$ (red-open symbols; left-hand scale) and $(1.5,1.5,2.5)_c$ (red-filled symbols; left-hand scale), in comparison to that of the reported paramagnetic scattering intensity (black symbols; right-hand scale) reported by Asai *et al.*[50].

## REFERENCES:

1 Ishikawa, A., Nohara, J. & Sugai, S. Raman Study of the Orbital-Phonon Coupling in ${\mathrm{L}\mathrm{a}\mathrm{C}\mathrm{o}\mathrm{O}}_{3}$. *Physical Review Letters* **93**, 136401, doi:10.1103/PhysRevLett.93.136401 (2004).

2 Ekanayaka, T. K., Maity, K. P., Doudin, B. & Dowben, P. A. Dynamics of Spin Crossover Molecular Complexes. *Nanomaterials* **12**, 1742 (2022).

3 Wu, Z. & Wentzcovitch, R. M. Spin crossover in ferropericlase and velocity heterogeneities in the lower mantle. *Proceedings of the National Academy of Sciences* **111**, 10468-10472, doi:doi:10.1073/pnas.1322427111 (2014).

4 Wu, Z. & Wentzcovitch, R. M. Elastic Anomalies in a Spin-Crossover System: Ferropericlase at Lower Mantle Conditions. *Physical Review Letters*, doi:10.1103/PhysRevLett.110.228501 (2013).

5 Shephard, G. E. Seismological Expression of the Iron Spin Crossover in Ferropericlase in the Earth's Lower Mantle. *Nat Commun*, doi:10.1038/s41467-021-26115-z (2021).

6 Holmström, E. & Stixrude, L. Spin Crossover in Ferropericlase from First-Principles Molecular Dynamics. *Physical Review Letters*, doi:10.1103/PhysRevLett.114.117202 (2015).

7 Ridier, K., Hoblos, A., Calvez, S., Lorenc, M., Nicolazzi, W., Cobo, S., Salmon, L., Routaboul, L., Molnár, G. & Bousseksou, A. Optical properties and photonic applications of molecular spin-crossover materials. *Coordination Chemistry Reviews* **535**, 216628, doi:https://doi.org/10.1016/j.ccr.2025.216628 (2025).

8 Senthil Kumar, K. & Ruben, M. Emerging trends in spin crossover (SCO) based functional materials and devices. *Coordination Chemistry Reviews* **346**, 176-205, doi:https://doi.org/10.1016/j.ccr.2017.03.024 (2017).

9 Zobel, C., Kriener, M., Bruns, D., Baier, J., Grüninger, M., Lorenz, T., Reutler, P. & Revcolevschi, A. Evidence for a low-spin to intermediate-spin state transition in ${\mathrm{LaCoO}}_{3}$. *Physical Review B* **66**, 020402, doi:10.1103/PhysRevB.66.020402 (2002).

10 Berggold, K., Kriener, M., Becker, P., Benomar, M., Reuther, M., Zobel, C. & Lorenz, T. Anomalous expansion and phonon damping due to the Co spin-state transition in $R{\text{CoO}}_{3}$ ($R=\text{La}$, Pr, Nd, and Eu). *Physical Review B* **78**, 134402 (2008).

11 Murata, S., Isida, S., Suzuki, M., Kobayashi, Y., Asai, K. & Kohn, K. Elastic anomalies with the spin-state transitions in LaCoO3. *Physica B: Condensed Matter* **263-264**, 647-649, doi:https://doi.org/10.1016/S0921-4526(98)01439-2 (1999).

12 Rotter, M., Wang, Z. S., Boothroyd, A. T., Prabhakaran, D., Tanaka, A. & Doerr, M. Mechanism of spin crossover in LaCoO3 resolved by shape magnetostriction in pulsed magnetic fields. *Scientific Reports* **4**, 7003, doi:10.1038/srep07003 (2014).

13 Jonker, G. H. & Santen, J. H. v. Magnetic compounds with perovskite structure III: ferromagnetic compounds of cobalt. *Physica* **19**, 120-130 (1953).

14 Goodenough, J. B. An interpretation of the magnetic properties of the perovskite-type mixed crystals $La_{1-x}Sr_xCoO_{3-\lambda}$. *J Phys Chem Solids* **6**, 287-297, doi:https://doi.org/10.1016/0022-3697(58)90107-0 (1958).

15 Haverkort, M. W., Hu, Z., Cezar, J. C., Burnus, T., Hartmann, H., Reuther, M., Zobel, C., Lorenz, T., Tanaka, A., Brookes, N. B., Hsieh, H. H., Lin, H. J., Chen, C. T. & Tjeng, L. H. Spin state transition in $LaCoO_3$ studied using soft x-ray absorption spectroscopy and magnetic circular dichroism. *Physical Review Letters* **97**, doi:10.1103/PhysRevLett.97.176405 (2006).

16 Tokura, Y., Okimoto, Y., Yamaguchi, S., Taniguchi, H., Kimura, T. & Takagi, H. Thermally induced insulator-metal transition in $LaCoO_3$: A view based on the Mott transition. *Phys. Rev. B* **58**, R1699-R1702, doi:10.1103/PhysRevB.58.R1699 (1998).

17 Takegami, D., Tanaka, A., Agrestini, S., Hu, Z., Weinen, J., Rotter, M., Schüßler-Langeheine, C., Willers, T., Koethe, T. C., Lorenz, T., Liao, Y. F., Tsuei, K. D., Lin, H. J., Chen, C. T. & Tjeng, L. H. Paramagnetic $LaCoO_3$: A Highly Inhomogeneous Mixed Spin-State System. *Physical Review X* **13**, 011037, doi:10.1103/PhysRevX.13.011037 (2023).

18 Feygenson, M., Novoselov, D., Pascarelli, S., Chernikov, R., Zaharko, O., Porcher, F., Karpinsky, D., Nikitin, A., Prabhakaran, D., Sazonov, A. & Sikolenko, V. Manifold of spin states and dynamical temperature effects in $LaCoO_3$: Experimental and theoretical insights. *Physical Review B* **100**, 054306, doi:10.1103/PhysRevB.100.054306 (2019).

19 Karolak, M., Izquierdo, M., Molodtsov, S. L. & Lichtenstein, A. I. Correlation-Driven Charge and Spin Fluctuations in $LaCoO_3$. *Physical Review Letters* **115**, 046401 (2015).

20 Shimizu, Y., Takahashi, T., Yamada, S., Shimokata, A., Jin-no, T. & Itoh, M. Symmetry Preservation and Critical Fluctuations in a Pseudospin Crossover Perovskite $LaCoO_3$. *Physical Review Letters* **119**, 267203 (2017).

21 Panfilov, A. S., Grechnev, G. E., Zhuravleva, I. P., Lyogenkaya, A. A., Pashchenko, V. A., Savenko, B. N., Novoselov, D., Prabhakaran, D. & Troyanchuk, I. O. Pressure effect on magnetic susceptibility of $LaCoO_3$. *Low Temperature Physics* **44**, 328-333, doi:10.1063/1.5030456 (2018).

22 Señarís-Rodríguez, M. A. & Goodenough, J. B. $LaCoO_3$ Revisited. *Journal of Solid State Chemistry* **116**, 224-231, doi:https://doi.org/10.1006/jssc.1995.1207 (1995).

23 Radaelli, P. G. & Cheong, S. W. Structural phenomena associated with the spin-state transition in $LaCoO_3$. *Physical Review B* **66**, 094408, doi:10.1103/PhysRevB.66.094408 (2002).

24 Podlesnyak, A., Streule, S., Mesot, J., Medarde, M., Pomjakushina, E., Conder, K., Tanaka, A., Haverkort, M. W. & Khomskii, D. I. Spin-State Transition in $LaCoO_3$: Direct Neutron Spectroscopic Evidence of Excited Magnetic States. *Physical Review Letters* **97**, 247208, doi:10.1103/PhysRevLett.97.247208 (2006).

25 Korotin, M. A., Ezhov, S. Y., Solovyev, I. V., Anisimov, V. I., Khomskii, D. I. & Sawatzky, G. A. Intermediate-spin state and properties of $LaCoO_3$. *Physical Review B* **54**, 5309-5316, doi:10.1103/PhysRevB.54.5309 (1996).

26 Phelan, D., Louca, D., Rosenkranz, S., Lee, S. H., Qiu, Y., Chupas, P. J., Osborn, R., Zheng, H., Mitchell, J. F., Copley, J. R. D., Sarrao, J. L. & Moritomo, Y. Nanomagnetic Droplets and Implications to Orbital Ordering in $La_{1-x}Sr_xCoO_3$. *Physical Review Letters* **96**, 027201, doi:10.1103/PhysRevLett.96.027201 (2006).

27 Maris, G., Ren, Y., Volotchaev, V., Zobel, C., Lorenz, T. & Palstra, T. T. M. Evidence for orbital ordering in $LaCoO_3$. *Physical Review B* **67**, 224423, doi:10.1103/PhysRevB.67.224423 (2003).

28 Kobayashi, Y., Naing, T. S., Suzuki, M., Akimitsu, M., Asai, K., Yamada, K., Akimitsu, J., Manuel, P., Tranquada, J. M. & Shirane, G. Inelastic neutron scattering study of phonons and magnetic excitations in $LaCoO_3$. *Physical Review B* **72**, 174405 (2005).

29 Doi, A., Fujioka, J., Fukuda, T., Tsutsui, S., Okuyama, D., Taguchi, Y., Arima, T., Baron, A. Q. R. & Tokura, Y. Multi-spin-state dynamics during insulator-metal crossover in $LaCoO_3$. *Physical Review B* **90**, 081109 (2014).

30 Baron, A. in *Synchrotron light sources and free electron lasers* (eds E. J. Jaeschke, S. Khan, J R Schneider, & J. B. Hastings) 1643-1758 (Springer, 2015).

31 Baron, A. Q. R., Uchiyama, H., Heid, R., Bohnen, K. P., Tanaka, Y., Tsutsui, S., Ishikawa, D., Lee, S. & Tajima, S. Two-phonon contributions to the inelastic x-ray scattering spectra of $MgB_2$. *Physical Review B* **75**, 020505, doi:10.1103/PhysRevB.75.020505 (2007).

32 Meyer, B., Elsässer, C., Lechermann, F. & Fähnle, M. *FORTRAN90 Program for Mixed-Basis Pseudopotential Calculations for Crystals* (Max-Planck-Institut für Metallforschung, Stuttgart).

33 Heid, R. & Bohnen, K.-P. Linear response in a density-functional mixed-basis approach. *Physical Review B* **60**, R3709-R3712 (1999).

34 Sugimoto, K., Ohsumi, H., Aoyagi, S., Nishibori, E., Moriyoshi, C., Kuroiwa, Y., Sawa, H. & Takata, M. Extremely High Resolution Single Crystal Diffractometory for Orbital Resolution using High Energy Synchrotron Radiation at SPring-8. *AIP Conference Proceedings* **1234**, 887-890, doi:10.1063/1.3463359 (2010).

35 al., H. M.-L. Z. e. HEiDi: Single crystal diffractometer at hot source. *Journal of large-scale research facilities* **1**, A7 (2015).

36 Ivashko, A., Manjo, T., Kauth, M., Tymoshenko, Y., Merritt, A., Bohnen, K.-P., Heid, R., Merz, M., Eich, A., Castellan, J.-P., Ivanov, A., Schreiber, N., Zheng, H., Mitchell, J. F., Meven, M., Park, J. T., Ishikawa, D., Nakamura, Y., Baron, A. Q. & Weber, F. Phonons reflect dynamic spin-state order in $LaCoO_3$. *KITopen data repository*, doi:https://doi.org/10.35097/xayn6n5pmcvdk776 (2026).

37 Baron, A. Q. R. Status of the RIKEN Quantum NanoDynamics Beamline (BL43LXU): The Next Generation for Inelastic X-Ray Scattering. *SPring-8 Inf. News* **15**, 14, doi:https://user.spring8.or.jp/sp8info/?p=3138. (2010).

38 Baron, A. Q. R. Introduction to High-Resolution Inelastic X-Ray Scattering. *arXiv:1504.01098* (2020).

39 Piovano, A. & Ivanov, A. The TAS-IN8 upgrade: Towards the limit of a three-axis spectrometer performance. *EPJ Web Conf.* **286**, 03011 (2023).

40 Gazizulina, A., Ivanov, A., Tymoshenko, Y. & Weber, F. in *ILL* (Institut Laue-Langevin (ILL), 2023).

41 Zentrum, H. M.-L. & al., e. PUMA: Thermal three axes spectrometer. *Journal of large-scale research facilities* **1**, A13, doi:https://doi.org/10.17815/jlsrf-1-36 (2015).

42 Kobayashi, Y., Mitsunaga, T., Fujinawa, G., Arii, T., Suetake, M., Asai, K. & Harada, J. Structural Phase Transition from Rhombohedral to Cubic in LaCoO 3. *Journal of the Physical Society of Japan* **69**, 3468-3469, doi:10.1143/JPSJ.69.3468 (2000).

43 Phelan, D. Constraints on the possible long-range orbital ordering in LaCoO3. *Journal of Magnetism and Magnetic Materials* **350**, 183-187, doi:https://doi.org/10.1016/j.jmmm.2013.09.001 (2014).

44 Fukuda, T., Mizuki, J., Ikeuchi, K., Yamada, K., Baron, A. Q. R. & Tsutsui, S. Doping dependence of softening in the bond-stretching phonon mode of $La_{2-x}Sr_xCuO_4$ ($0 \leq x \leq 0.29$). *Physical Review B* **71**, 060501, doi:10.1103/PhysRevB.71.060501 (2005).

45 Reznik, D., Pintschovius, L., Ito, M., Iikubo, S., Sato, M., Goka, H., Fujita, M., Yamada, K., Gu, G. D. & Tranquada, J. M. Electron-phonon coupling reflecting dynamic charge inhomogeneity in copper oxide superconductors. *Nature* **440**, 1170 (2006).

46 d'Astuto, M., Dhalenne, G., Graf, J., Hoesch, M., Giura, P., Krisch, M., Berthet, P., Lanzara, A. & Shukla, A. Sharp optical-phonon softening near optimal doping in $La_{2-x}Ba_xCuO_{4+\delta}$ observed via inelastic x-ray scattering. *Physical Review B* **78**, 140511 (2008).

47 Wang, Q., von Arx, K., Horio, M., Mukkattukavil, D. J., Küspert, J., Sassa, Y., Schmitt, T., Nag, A., Pyon, S., Takayama, T., Takagi, H., Garcia-Fernandez, M., Zhou, K.-J. & Chang, J. Charge order lock-in by electron-phonon coupling in $La_{1.675}Eu_{0.2}Sr_{0.125}CuO_4$. *Science Advances* **7**, eabg7394, doi:10.1126/sciadv.abg7394 (2021).

48 Sun, Z., Chuang, Y.-D., Fedorov, A. V., Douglas, J. F., Reznik, D., Weber, F., Aliouane, N., Argyriou, D. N., Zheng, H., Mitchell, J. F., Kimura, T., Tokura, Y., Revcolevschi, A. & Dessau, D. S. Quasiparticlelike peaks, kinks, and electron-phonon coupling in the (π,0) regions in the CMR oxide $La_{2-2x}Sr_{1+2x}Mn_2O_7$. *Physical Review Letters* **97**, 056401 (2006).

49 Weber, F., Aliouane, N., Zheng, H., Mitchell, J. F., Argyriou, D. N. & Reznik, D. Signature of checkerboard fluctuations in the phonon spectra of a possible polaronic metal $La_{1.2}Sr_{1.8}Mn_2O_7$. *Nature materials* **8**, 798-802 (2009).

50 Asai, K., Yokokura, O., Nishimori, N., Chou, H., Tranquada, J. M., Shirane, G., Higuchi, S., Okajima, Y. & Kohn, K. Neutron-scattering study of the spin-state transition and magnetic correlations in $La_{1-x}Sr_xCoO_3$ (x=0 and 0.08). *Physical Review B* **50**, 3025-3032, doi:10.1103/PhysRevB.50.3025 (1994).

51 Ishikawa, A., Nohara, J. & Sugai, S. Raman Study of the Orbital-Phonon Coupling in $LaCoO_3$. *Phys. Rev. Lett.* **93**, 136401, doi:10.1103/PhysRevLett.93.136401 (2004).

52 Gnezdilov, V., Fomin, V., Yeremenko, A. V., Choi, K. Y., Pashkevich, Y., Lemmens, P., Shiryaev, S., Bychkov, G. & Barilo, S. Low-temperature mixed spin state of Co3+ in LaCoO3 evidenced from Jahn-Teller lattice distortions. *Low Temperature Physics* **32**, 162-168, doi:10.1063/1.2171521 (2006).

53 Sikolenko, V. V., Molodtsov, S. L., Izquierdo, M., Troyanchuk, I. O., Karpinsky, D., Tiutiunnikov, S. I., Efimova, E., Prabhakaran, D., Novoselov, D. & Efimov, V. Correlated oxygen displacements and phonon mode changes in LaCoO3 single crystal. *Physica B: Condensed Matter* **536**, 597-599, doi:https://doi.org/10.1016/j.physb.2017.10.031 (2018).

54 Krannich, S., Sidis, Y., Lamago, D., Heid, R., Mignot, J. M., Lohneysen, H. v., Ivanov, A., Steffens, P., Keller, T., Wang, L., Goering, E. & Weber, F. Magnetic moments induce strong phonon renormalization in FeSi. *Nat Commun* **6**, 8961, doi:https://doi.org/10.1038/ncomms9961 (2015).

55 Raccah, P. M. & Goodenough, J. B. First-Order Localized-Electron to Collective-Electron Transition in $LaCoO_3$. *Physical Review* **155**, 932-943, doi:10.1103/PhysRev.155.932 (1967).

56 Merz, M., Nagel, P., Pinta, C., Samartsev, A., v. Löhneysen, H., Wissinger, M., Uebe, S., Assmann, A., Fuchs, D. & Schuppler, S. X-ray absorption and magnetic circular dichroism of $LaCoO_3$, $La_{0.7}Ce_{0.3}CoO_3$, and $La_{0.7}Sr_{0.3}CoO_3$ films: Evidence for cobalt-valence-dependent magnetism. *Physical Review B* **82**, 174416, doi:10.1103/PhysRevB.82.174416 (2010).

57 Maschek, M., Castellan, J. P., Lamago, D., Reznik, D. & Weber, F. Polaronic correlations and phonon renormalization in $La_{1-x}Sr_xMnO_3$ ($x$ = 0.2, 0.3). *Physical Review B* **97**, 245139, doi:10.1103/PhysRevB.97.245139 (2018).

58 Maschek, M., Lamago, D., Castellan, J. P., Bosak, A., Reznik, D. & Weber, F. Polaronic metal phases in $La_{0.7}Sr_{0.3}MnO_3$ uncovered by inelastic neutron and x-ray scattering. *Physical Review B* **93**, 045112 (2016).